\documentclass[iop]{emulateapj}

\usepackage{natbib}
\usepackage{ulem}
\usepackage{comment}
\usepackage{txfonts}

\usepackage{color} 

\usepackage[T1]{fontenc}

\usepackage{bm}

\newcommand{\eg}{e.g., }

\newcommand{\Msun}{\ensuremath{M_{\odot}}}
\newcommand{\Rsun}{\ensuremath{R_{\odot}}}
\newcommand{\kms}{km~s$^{-1}$ }

\newcommand{\magdays}{mag~day$^{-1}$}

\newcommand{\Nifs}{$^{56}$Ni~}

\def\gsim{\mathrel{\rlap{\lower 4pt \hbox{\hskip 1pt $\sim$}}\raise 1pt
\hbox {$>$}}}
\def\lsim{\mathrel{\rlap{\lower 4pt \hbox{\hskip 1pt $\sim$}}\raise 1pt
\hbox {$<$}}}

\newcommand{\si}{{\sc i}~}
\newcommand{\sii}{{\sc ii}~}
\newcommand{\siii}{{\sc iii}~}

\newcommand{\Ha}{H$\alpha$ }
\newcommand{\Hb}{H$\beta$ }
\newcommand{\hs}{\hspace{-1.5mm}}
\newcommand{\habs}{\hspace{-1mm}}
\newcommand{\bvri}{{\it BVRI}-band }
\newcommand{\jhk}{{\it JHK$_{s}$}-band }

\newcommand{\red}{\textcolor{black}}

\shorttitle{Rapidly Evolving Supernova SN~2017czd}
\shortauthors{Nakaoka T., et al.}

\begin{document}

\title{
SN~2017\MakeLowercase{czd}: A rapidly evolving supernova from a weak explosion of a type II\MakeLowercase{b} supernova progenitor
}
\author{
  Tatsuya Nakaoka\altaffilmark{1},
  Takashi J. Moriya\altaffilmark{2},
  Masaomi Tanaka\altaffilmark{3},
  Masayuki Yamanaka\altaffilmark{4},
  Koji S. Kawabata\altaffilmark{1,4},
  Keiichi Maeda\altaffilmark{5},
  Miho Kawabata\altaffilmark{1},
  Naoki Kawahara\altaffilmark{1},
  Koichi Itagaki\altaffilmark{6},
  Ryoma Ouchi\altaffilmark{5},
  Sergei I. Blinnikov\altaffilmark{7,8,9},
  Nozomu Tominaga\altaffilmark{10}, and
  Makoto Uemura\altaffilmark{1,4}.
}

\altaffiltext{1}{Department of Physical Science, Hiroshima University, Kagamiyama 1-3-1, Higashi-Hiroshima 739-8526, Japan;nakaoka@astro.hiroshima-u.ac.jp}
\altaffiltext{2}{National Astronomical Observatory of Japan, 2-21-1 Osawa, Mitaka, Tokyo 181-8588, Japan}
\altaffiltext{3}{Astronomical Institute, Graduate School of Science, Tohoku University, Aramaki, Aoba, Sendai 980-8578, Japan}
\altaffiltext{4}{Hiroshima Astrophysical Science Center, Hiroshima University, Higashi-Hiroshima, Hiroshima 739-8526, Japan}
\altaffiltext{5}{Department of Astronomy, Kyoto University,Kitashirakawa-Oiwake-cho, Sakyo-ku, Kyoto 606-8502, Japan}
\altaffiltext{6}{Itagaki Astronomical Observatory, Yamagata 990-2492, Japan}
\altaffiltext{7}{Institute for Theoretical and Experimental Physics (ITEP), 117218 Moscow, Russia}
\altaffiltext{8}{Space Research Institute (IKI), RAS, 117997 Moscow, Russia}
\altaffiltext{9}{Kavli Institute for the Physics and Mathematics of the Universe (WPI), The University of Tokyo Institutes for Advanced Study, The University of Tokyo, 5-1-5 Kashiwanoha, Kashiwa, Chiba 277-8583, Japan}
\altaffiltext{10}{Department of Physics, Faculty of Science and Engineering, Konan University, Kobe, Hyogo 658-8501, Japan}

\begin{abstract}

We present optical and near-infrared observations of the rapidly evolving supernova (SN) 2017czd that shows hydrogen features. 
The optical light curves exhibit a short plateau phase ($\sim \habs 13$~days in the $R$-band) followed by a rapid decline by $4.5$ mag in $\sim 20~\mathrm{days}$ after the plateau.
The decline rate is larger than those of any standard SNe,
and close to those of rapidly evolving transients. 
The peak absolute magnitude is $-16.8$ mag in the $V$-band, which is within the observed range for SNe IIP and rapidly evolving transients.
The spectra of SN~2017czd clearly show the hydrogen features and resemble those of SNe IIP at first.
The \Ha line, however, does not evolve much with time and it becomes similar to those in SNe IIb at decline phase.
We calculate the synthetic light curves using a SN IIb progenitor which has  16~$\Msun$ at the zero-age main sequence and evolves in a binary system.
The model with a low explosion energy ($5\times 10^{50}$~erg) and a low \Nifs mass ($0.003~\Msun$) can reproduce the short plateau phase as well as the sudden drop of the light curve as observed in SN~2017czd. 
We conclude that SN~2017czd might be the first identified weak explosion from a SN IIb progenitor.
We suggest that some rapidly evolving transients can be explained by such a weak explosion of the progenitors with little hydrogen-rich envelope.

\end{abstract}


\section{Introduction}
\label{intro}

 Supernovae (SNe) are classified into some classes using the spectral features and light curve properties,
which reflect the diversity in their progenitor stars and explosion mechanisms. 
SNe with hydrogen absorption features in their spectra and the plateau in their optical light curves are classified as Type II-plateau SNe \citep[SNe IIP; ][]{filippenko1997}. 
Through the analysis of pre-explosion data  \citep{smartt2009,smartt2015}, red supergiants are identified as progenitors of SNe IIP in full accord with standard theoretical predictions \citep[e.g.,][]{grassberg1971,2003ApJ...591..288H}.
SNe IIb exhibit both hydrogen and helium lines in their early spectra,
and their optical light curves show one or two peaks \citep{bersten2018}.
Yellow supergiant stars have been detected in pre-explosion images for some SNe IIb \citep[e.g.,][]{Aldering1994,Crockett2008,VanDyk2014}. 
SNe Ib/c do not show hydrogen features \citep{hunter2009,benetti2011,takaki2013}.
Although the progenitor detection is still limited for SNe Ib/c \citep{cao2013,folatelli2016,2016MNRAS.461L.117E,vandyk2018}, 
the progenitors should be massive stars in which hydrogen-rich envelopes are stripped either by stellar wind or binary interaction.

Through the analysis of light curves and spectra, combined with that of pre-explosion images,
relationships between explosion and progenitor properties have been well studied for major classes of SNe
\citep[e.g.,][]{2003ApJ...582..905H,2015ApJ...806..225P,utrobin2015,lyman2016}.
However, the progenitor stars are still unclear for some peculiar SNe \citep[e.g.,][]{nakaoka2018,jaeger2018,arcavi2017}.
One example is so-called rapidly evolving transients such as those discovered by Panoramic Survey Telescope and Rapid Response System (Pan-STARRS) \citep{drout2014}.
Their rise and decline rates are much faster
than fast-evolving SNe Ib/c like SN 1994I \citep{richmond1996_94I}.
Recently, an increasing number of rapid transients have been found with wide-field and high-cadence survey,
\eg Subaru Hyper Suprime-Cam Transient Survey \citep{tanaka2016}, Palomar Transient Factory \citep{2017ApJ...851..107W}, Dark Energy Survey Supernova Program \citep{pursiainen2018}, and K2/Kepler \citep{rest2018}. 
Most of them are discovered in star-forming galaxies, implying massive progenitors. 
However, the exact nature of the progenitors and explosions remain elusive.


The absolute peak magnitudes of rapidly evolving transients have a large variety, $-15 - -22$ mag \citep{drout2014,pursiainen2018}. 
Spectral features also show a diversity:
some of them show the blue featureless spectra with photospheric temperatures of 20000-30000 K 
\citep{drout2014,pursiainen2018}
while others show the absorption features including helium lines, e.g., for SNe 2002bj \citep{poznanski2010}, 2005ek \citep{drout2013} and 2010X \citep{kasliwal2010}. 
However, hydrogen features have never been clearly identified in rapidly evolving transients.

Many different kinds of scenarios have been proposed for the rapidly evolving transients. For example, rapidly evolving transients may be related to peculiar core-collapse SNe, \eg ultra-stripped SNe \citep[e.g.,][]{tauris2013, moriya2017_us,De201}, electron-capture SNe \citep[e.g.,][]{moriya2016}, magnetar-powered SNe \citep[e.g.,][]{2017ApJ...851..107W}, and failed core-collapse SN explosions \citep[e.g.,][]{moriya2010}. The failed core-collapse SN explosions may also lead to the fallback accretion-powered rapidly evolving SNe \citep[e.g.,][]{2013ApJ...772...30D}.
Some others may be powered by the interaction between the SN ejecta and the circumstellar material (CSM) \citep[e.g.,][]{2018MNRAS.475.3152K},
which in fact nicely reproduces the light curves of well-observed KSN~2015K \citep{rest2018}. 
Other proposed mechanisms include 
tidal disruption events \citep{drout2014}
and peculiar thermonuclear explosions \citep{bildsten2007,shen2010}.
In fact, the natures of the progenitor and explosion are not clear for many rapidly evolving transients partly due to the lack of good photometric and spectroscopic coverage. 
To understand the nature of rapid transients, extensive follow-up observations are necessary.


In this paper, we present our observations of rapidly evolving SN 2017czd.
SN~2017czd was discovered by Koichi Itagaki on 2017 April 12.7 (UT) \citep{itagaki2017}.
About 1 day after the discovery, this SN was classified as a young SN II with the featureless and blue continuum through the spectroscopic observation \citep{hosse2017}.
The host galaxy, UGC~9567, is an irregular galaxy with emission lines.
The date of explosion is constrained by the last non-detection date (Apr 10.3) by Gaia Photometric Alerts. 
In this paper, the explosion date is assumed to be 2017 Apr 11.5 (defined as $t=0$~days, where $t$ is the time since the explosion in the rest frame), which is the middle of the last non detection and the first detection.
The observational epochs are given with respect to the explosion date.

This paper is organized as follows.
We introduce our observations and the data reduction in \S 2. We show the observed light curves and compare them with other SNe and other rapidly evolving transients in \S 3. 
We present our spectra and discuss similarities with SNe II in \S 4. 
In \S 5, we discuss the nature of SN 2017czd by constructing the bolometric light curve and by comparing it with the results of radiative transfer calculations based on a binary progenitor model.
Finally, we give summary in \S 6. 
Throughout of the paper, the distance to the host galaxy is assumed to be 32.0 $\pm$ 1.5 Mpc, adopting the distance from the recession velocity ($z=0.009$; via NED\footnote{http://ned.ipac.caltech.edu/}).

\begin{figure}[t]
\centering
\includegraphics[width=6cm,clip]{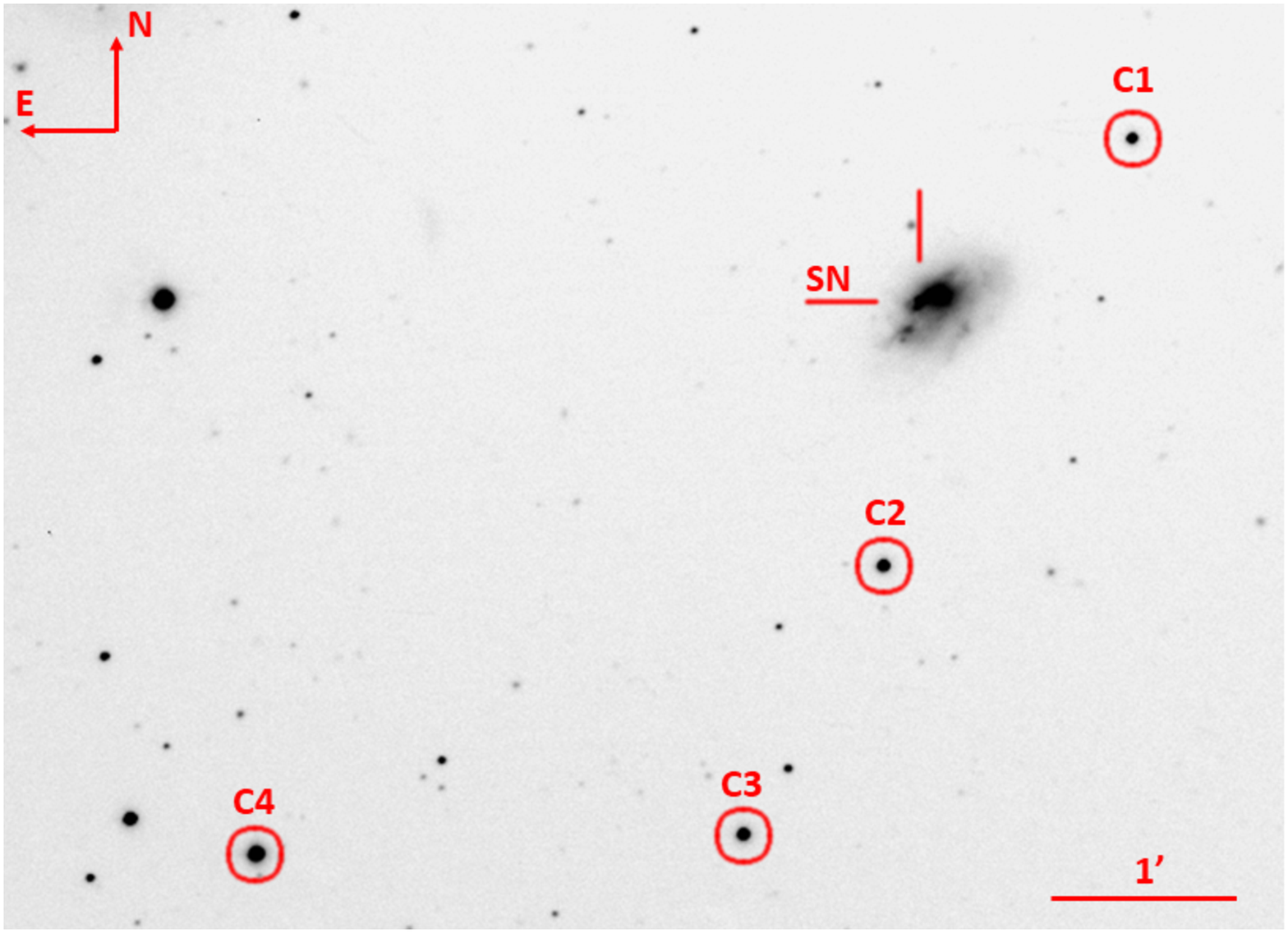}
\vspace{0mm}
\includegraphics[width=6cm,clip]{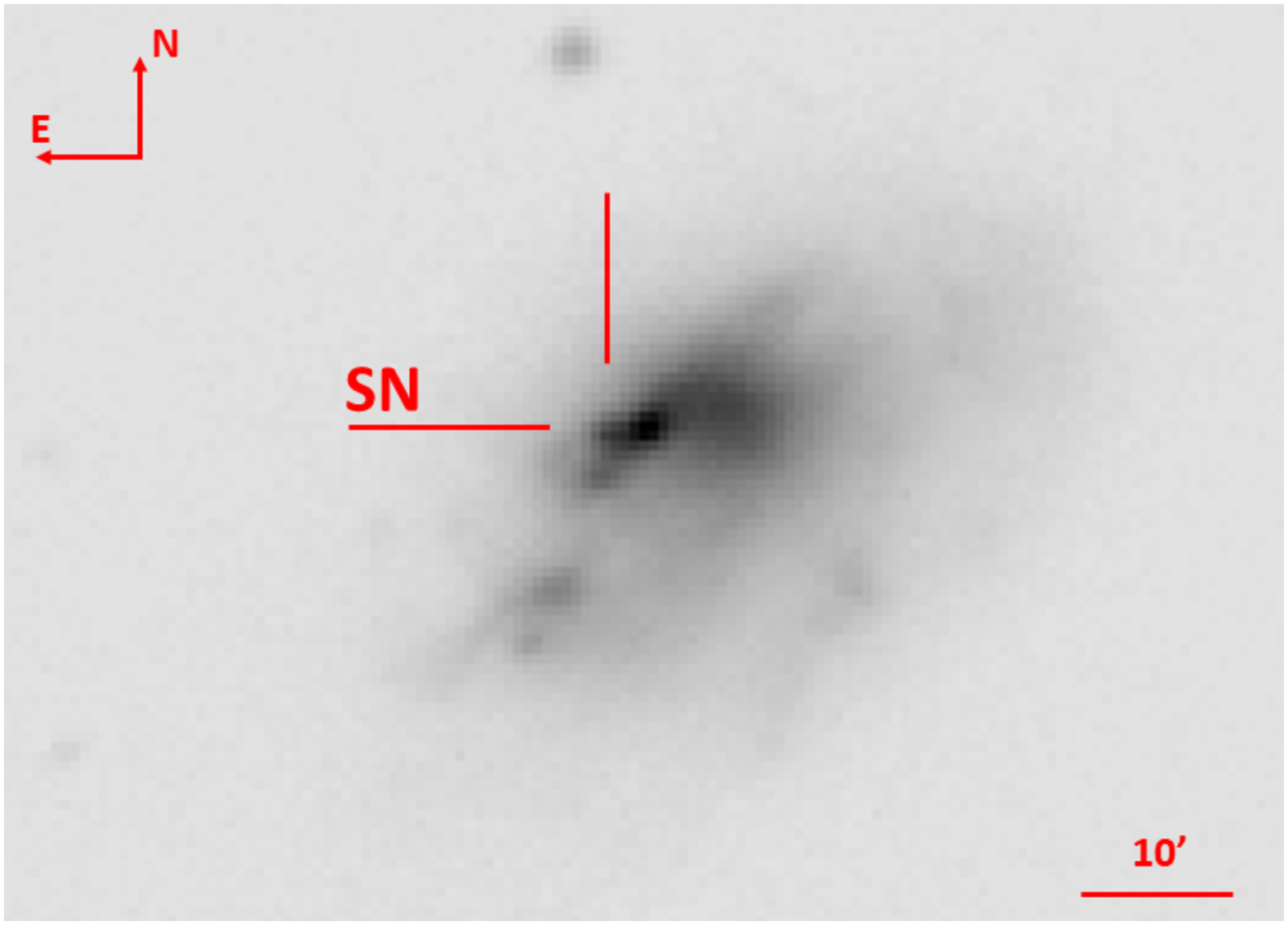}
\includegraphics[width=6cm,clip]{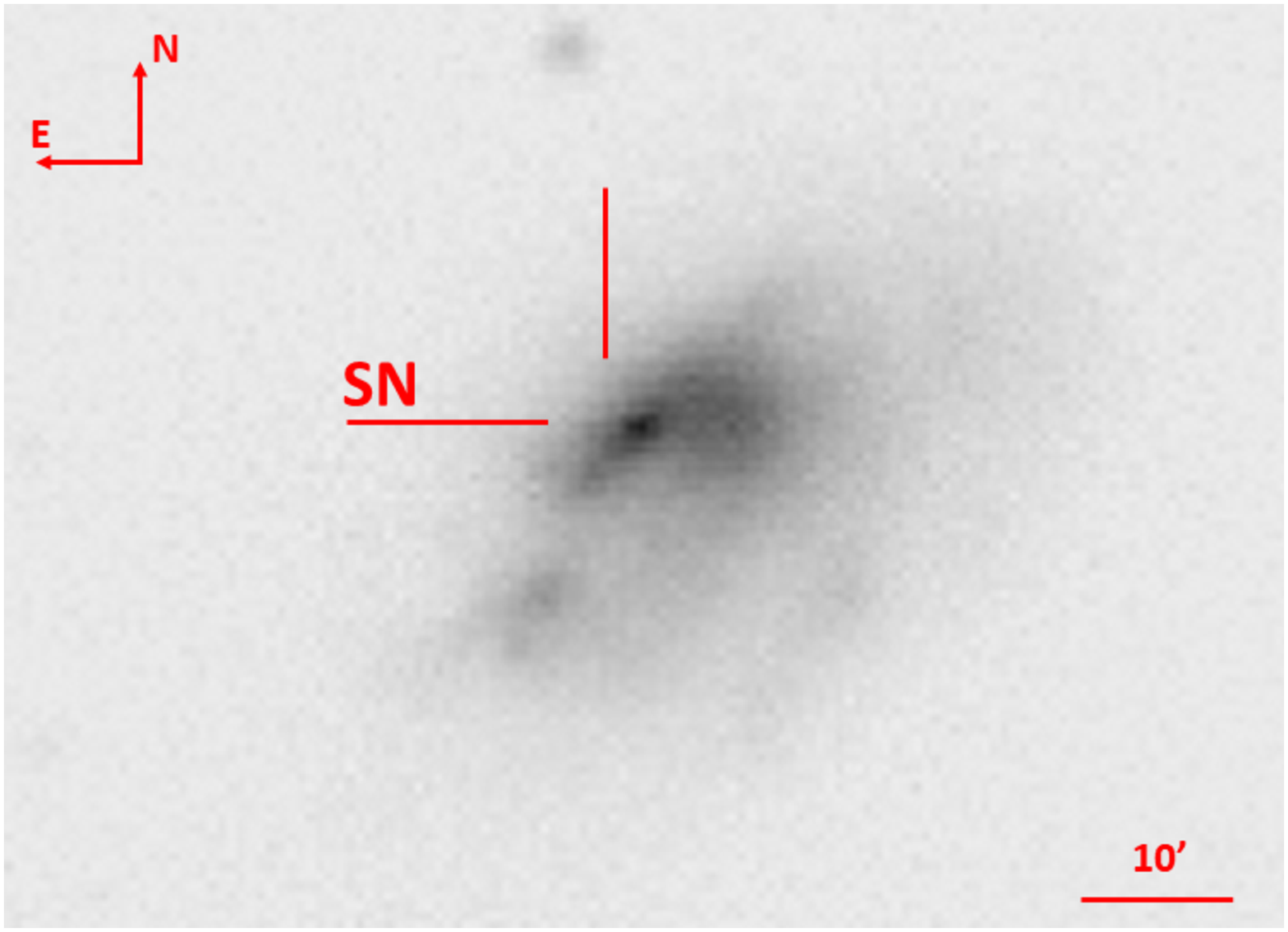}
\caption{(Top) The $R$-band image of SN 2017czd in UGC~9567 taken at $t = 5.0$ days using HOWPol. The location of the SN is marked by the two lines. Comparison stars are marked by the circles.
(Middle) The same as the top panel but closer to the SN position.
(Bottom) The same as the middle panel but at $t = 327$~days.}
\label{fig:fc}
\end{figure}

\begin{deluxetable}{lllll}
\tablewidth{0pt}
\tablecaption{Optical magnitudes of the comparison stars}
\tablehead{
  ID &  B  & V & R & I \\
  &  (mag) & (mag) & (mag) & (mag)
}
\startdata
C1 & 17.75(0.03) & 17.15(0.02) & 16.67(0.01) & 16.2(0.06)\\
C2 & 17.06(0.03) & 16.56(0.02) & 16.11(0.02) & 15.71(0.04)\\
C3 & 16.62(0.04) & 16.28(0.06) & 15.94(0.02) & 15.57(0.03)\\
C4 & 15.99(0.04) & 15.48(0.01) & 14.83(0.05) & 14.36(0.04)
\enddata
\label{table:com}
\end{deluxetable}

\section{Observations and Data Reduction}

Optical imaging data were obtained
by using the Hiroshima One-shot Wide-field Polarimeter \citep[HOWPol;][]{kawabata2008}
and the Hiroshima Optical and Near-InfraRed Camera \citep[HONIR;][]{Sakimoto2012,akitaya2014,Ui2014} 
installed to the 1.5-m Kanata telescope
at the Higashi-Hiroshima Observatory, Hiroshima University.
We obtained \bvri data with HOWPol in 30 nights from 2017 April 14.6 ($t=3.1$ days) to 2018 April 18.5 ($t = 327$~days),
and HONIR in 18 nights from 2017 April 13.6 ($t=2.1$ days) to 2018 April 18.7 ($t = 327$~days).
The last images obtained by HOWPol and HONIR are used for subtraction from the host galaxy.
All magnitudes are given in the Vega magnitudes throughout the paper.

\begin{figure}[t]
\centering
\includegraphics[width=8cm,clip]{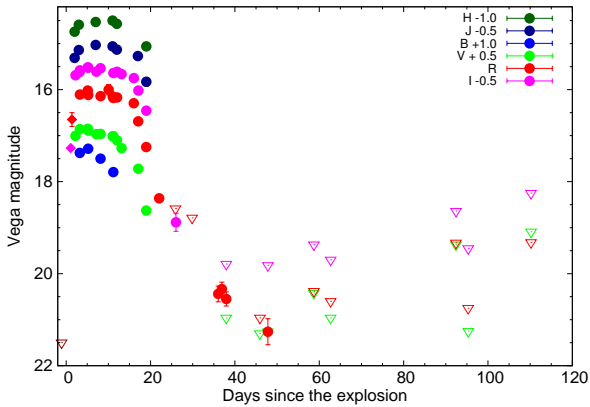}
\caption{Optical and NIR light curves of SN~2017czd.
Circles show the HOWPol and HONIR data. The diamond shapes at the early epochs show non-filter magnitude in K. Itagaki's image (red) and Pan-STARRS $i$-band magnitude (magenta). The Galactic extinction \citep{schlafly2011} has been corrected.
The upper limits are denoted by the triangles.}
\label{fig:lc}
\end{figure}

Contamination from the underlying emission from the host galaxy cannot be ignored in the optical data (see Figure \ref{fig:fc}).
For optical photometry, we first performed image subtraction using the 
HOWPol and HONIR data taken at 2018 Apr 18 ($t = 327$~days).
Then, we measured the brightness in the subtracted images using aperture photometry in the {\it IRAF/DAOPHOT} package \citep{stetson1987}. 
Local comparison star magnitudes were calibrated 
using the photometric standard stars \citep{landolt1992} (see Figure \ref{fig:fc} and Table \ref{table:com}).
The color terms were also corrected.
The obtained magnitudes are summarized in Table \ref{table:opt}.
Figure \ref{fig:lc} shows the light curves of SN~2017czd. 

We also performed the point-spread function (PSF) photometry of the SN in 
the discovery image obtained by Itagaki.
We regard the non-filter magnitude as $R$-band from the literature \citep{Zheng2014}.
This photometric information is also included in the $R$ band light curves presented in this paper.

For the part of the $VRI$-band images obtained using HOWPol and HONIR after $t=25.7$~days,
the SN was not detected. We derived 5 sigma upper-limit magnitudes by measuring dispersion of 
the background sky brightness using the same apertures. 
The light curves show these upper limits (Figure \ref{fig:lc}).

We also obtained the optical images of SN 2017czd using the Gemini
Multi-Object Spectrograph \citep[GMOS,][]{hook2004}
attached to the Gemini telescope on 2018 Jul 15 ($t = 456$~days).
We confirmed that there was no object brighter than the underlying component of the host galaxy at the  position of the SN in the $r$-band images.
We also confirmed that the brightness of the underlying component of the host galaxy
is consistent with pre-explosion images obtained by Sloan Digital Sky Survey (SDSS) and Pan-STARRS.

The near infrared (NIR) imaging data were obtained using HONIR for 16 nights from 2017 Apr 13.6 ($t = 2.1$ days) to May 27.6 ($t = 45.8$~days).
We took images with dithering to accurately subtract the bright foreground sky.
We did not subtract the underlying component of the host galaxy for the NIR imaging data since its contamination was less than $\sim10\%$ ($t=2.1-18.8$~days).
After the standard data reduction, we carried out the PSF photometry.
Photometric calibrations were performed using the magnitudes of reference
stars given in the 2MASS catalog \citep{persson1998}.
The derived \jhk magnitudes and the light curves are shown in Table
\ref{table:nir} and Figure \ref{fig:lc}, respectively.

In the light curves, the Galactic extinction of $E(B-V)=0.02$ \citep{schlafly2011} has been corrected. 
Since the absorption lines of Na~\si~D at the wavelength corresponding to the host galaxy are not detected (equivalent width $\lesssim \hs 0.062$ \AA), 
we assume that the extinction within the host galaxy is negligible. 

\begin{deluxetable*}{lllllll}
\tablewidth{0pt}
\tablecaption{Log of the optical photometry of SN~2017czd.}
\tablehead{
  MJD &  Epoch     & B & V & R & I & Instrument\\
  &  (day)  & (mag) & (mag) & (mag) & (mag) & 
}
\startdata
57856.6 & 2.1 & --- & 16.5(0.07) & --- & 16.19(0.05) & HONIR\\
57857.6 & 3.1 & 16.38(0.03) & 16.36(0.02) & 16.11(0.02) & 16.08(0.02) & HONIR\\
57859.6 & 5.0 & 16.28(0.02) & 16.39(0.03) & 16.12(0.04) & 16.02(0.03) & HOWPol\\
57861.6 & 7.0 & --- & 16.47(0.06) & --- & 16.11(0.05) & HONIR\\
57862.6 & 8.0 & 16.5(0.05) & 16.46(0.02) & 16.14(0.05) & 16.04(0.02) & HOWPol\\
57864.5 & 9.9 & --- & --- & 16.02(0.11) & --- & HOWPol\\
57865.7 & 11.1 & 16.79(0.02) & 16.51(0.02) & 16.18(0.06) & 16.14(0.02) & HOWPol\\
57866.6 & 12.0 & --- & 16.6(0.04) & 16.19(0.04) & 16.11(0.04) & HONIR\\
57867.6 & 13.0 & --- & 16.77(0.05) & --- & 16.16(0.04) & HOWPol\\
57870.6 & 15.9 & --- & --- & 16.32(0.05) & 16.25(0.05) & HOWPol\\
57871.6 & 16.9 & --- & 17.22(0.07) & 16.71(0.05) & 16.52(0.04) & HONIR\\
57873.5 & 18.8 & --- & 18.13(0.07) & 17.27(0.05) & 16.96(0.05) & HONIR\\
57876.6 & 21.9 & --- & --- & 18.38(0.06) & --- & HOWPol\\
57880.5 & 25.7 & --- & --- & >18.59 & 19.38(0.19) & HONIR\\
57884.5 & 29.7 & --- & --- & >18.79 & --- & HONIR\\
57890.8 & 36.0 & --- & --- & 20.44(0.17) & --- & HOWPol\\
57891.6 & 36.7 & --- & --- & 20.34(0.15) & --- & HOWPol\\
57892.7 & 37.8 & --- & --- & 20.55(0.15) & >20.29 & HOWPol\\
57900.7 & 45.8 & --- & >20.8 & >20.96 & --- & HOWPol\\
57902.6 & 47.7 & --- & >20.76 & 21.26(0.28) & >20.32 & HOWPol\\
57913.6 & 58.6 & --- & >19.92 & >20.38 & >19.87 & HOWPol\\
57917.6 & 62.5 & --- & >20.46 & >20.6 & >20.2 & HOWPol\\
57947.6 & 92.2 & --- & >18.87 & >19.33 & >19.14 & HOWPol\\
57950.6 & 95.2 & --- & >20.75 & >20.75 & >19.95 & HOWPol

\enddata
\label{table:opt}
\end{deluxetable*}

\begin{deluxetable}{lllll}
\tablewidth{0pt}
\tablecaption{Log of the NIR photometry of SN~2017czd.}
\tablehead{
  MJD &  Epoch     & J & H\\
  &  (day)     & (mag) & (mag)
}
\startdata
57856.6 & 2.1 & 15.81(0.03) & 15.75(0.03)\\
57857.6 & 3.1 & 15.64(0.03) & 15.59(0.03)\\
57861.6 & 7.0 & 15.54(0.03) & 15.53(0.03)\\
57865.7 & 11.1 & 15.56(0.02) & 15.51(0.03)\\
57866.6 & 12.0 & 15.64(0.03) & 15.58(0.04)\\
57871.6 & 16.9 & 15.77(0.03) & ---\\
57873.5 & 18.8 & 16.33(0.09) & 16.06(0.05)
\enddata
\label{table:nir}
\end{deluxetable}

\begin{figure}[t]
\centering
\includegraphics[width=8cm,clip]{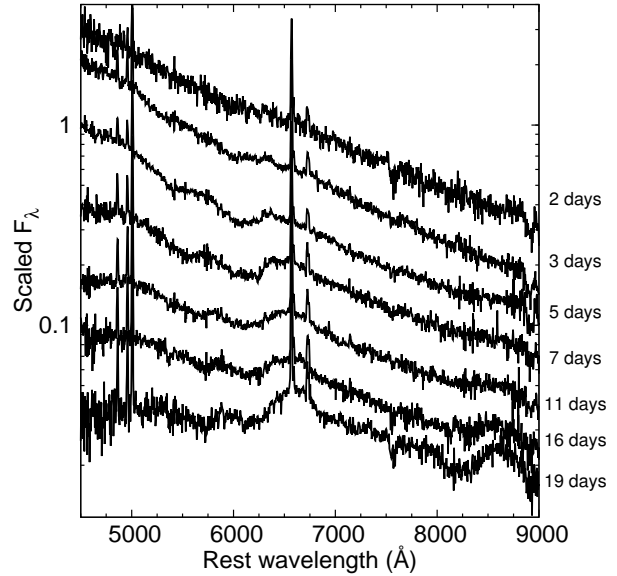}
\caption{Spectral evolution of SN~2017czd.
The epoch for each spectrum is given in days since the explosion.
}
\label{fig:spec}
\end{figure}

We also performed optical spectroscopic observations using HOWPol
for 9 nights from 2017 Apr 13.6 ($t=2.1$ days) to 30.7 ($t=19.0$~days).
We used a grism with a spectral resolution of $R \sim 400$
and a spectral coverage of 4500--9000~\AA.
We observed spectroscopic standard stars in the same nights for the flux calibration.
For the wavelength calibration, we used the sky emission lines in object frames.
The strong atmospheric absorption bands around 6900 and 7600 \AA\ have
been removed using smooth spectra of the hot standard stars.
Additionally, the continuum of the spectra is modified to match the SED from the photometry at the same epoch if necessary.
The log of our spectroscopic observations is given in Table \ref{table:spec},
and the obtained spectra are shown in Figure \ref{fig:spec}.

\begin{deluxetable}{lll} 
\tablewidth{0pt}
\tablecaption{Log of the spectroscopic observations of SN~2017czd.}
\tablehead{
  MJD  &  Epoch     & Exposure \\
  & (day) & (sec)
}
\startdata
57856.6 & 2.1 & 2400\\
57857.6 & 3.1 & 2700\\
57859.6 & 5.0 & 2700\\
57861.8 & 7.2 & 2700\\
57865.6 & 11.0 & 2700\\
57866.6 & 12.0 & 2700\\
57867.7 & 13.1 & 3600\\
57870.7 & 16.0 & 2700\\
57873.7 & 19.0 & 2700

\enddata
\label{table:spec}
\end{deluxetable}

\section{Light Curves}\label{sec:lightcurve}

\subsection{Light Curve Properties}

\begin{figure}[t]
\includegraphics[width=8cm,clip]{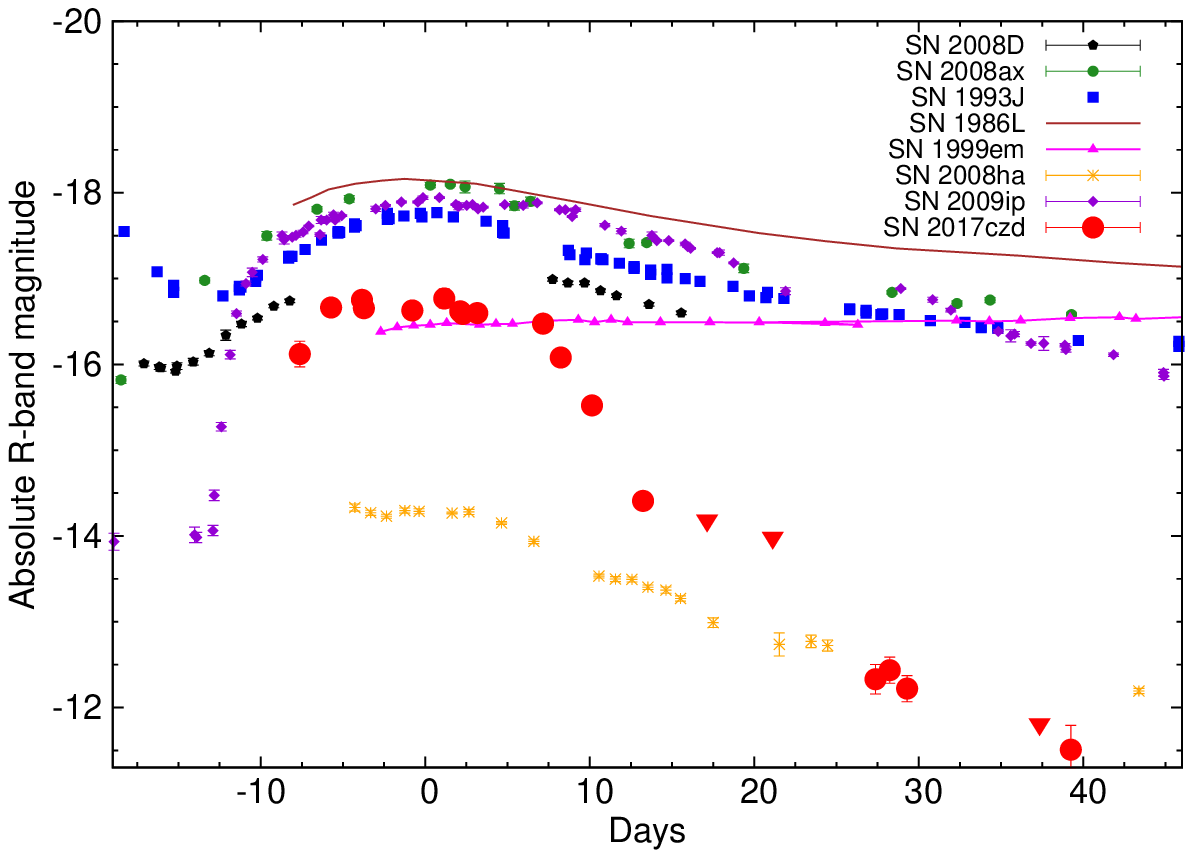}
\includegraphics[width=8cm,clip]{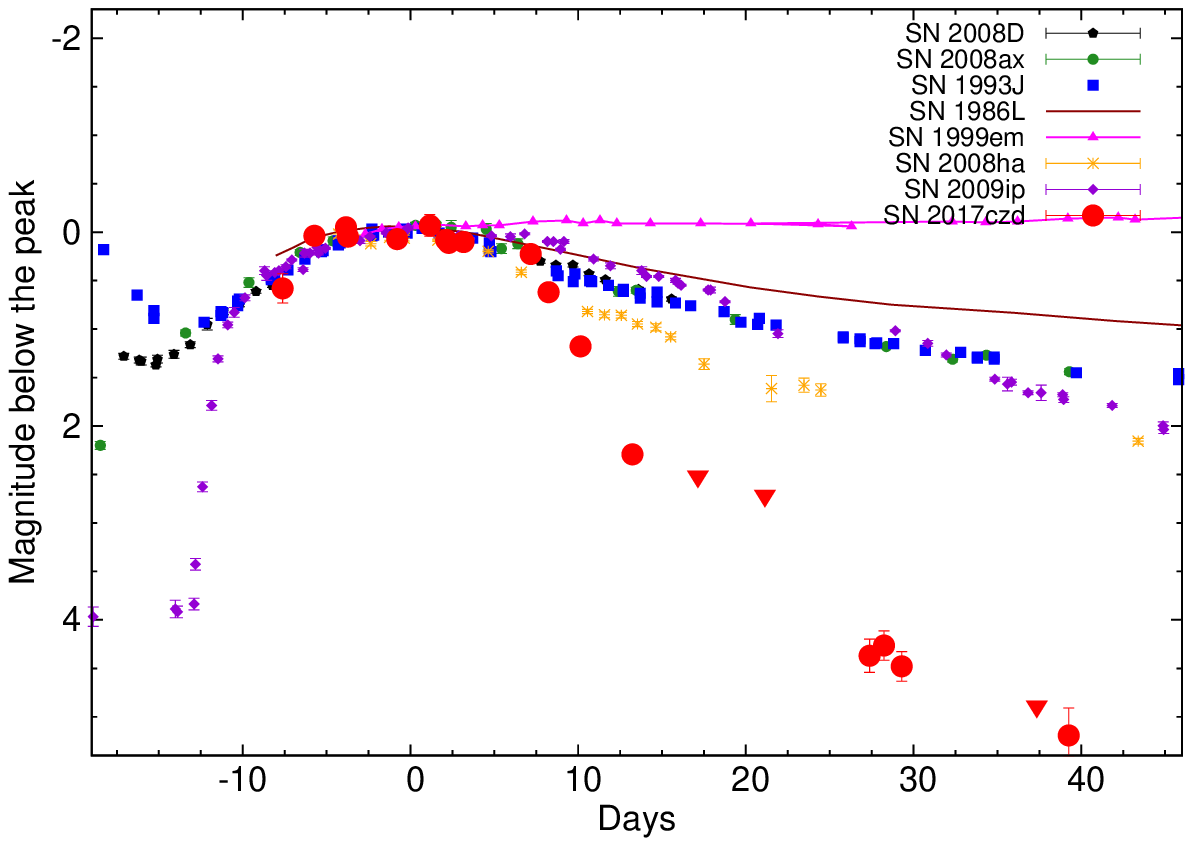}
\caption{(Top) The $R$-band absolute magnitudes of SN~2017czd compared with those of well-observed SNe. The extinction of each SNe has been corrected.
Light curves are shifted in the time axis to match their maximum dates.
(Bottom) The same as the top panel but the magnitudes  relative to the peaks.
Both figures are compared in the rest frames.}
\label{fig:abs_normal}
\end{figure}

\begin{figure}[t]
\includegraphics[width=8cm,clip]{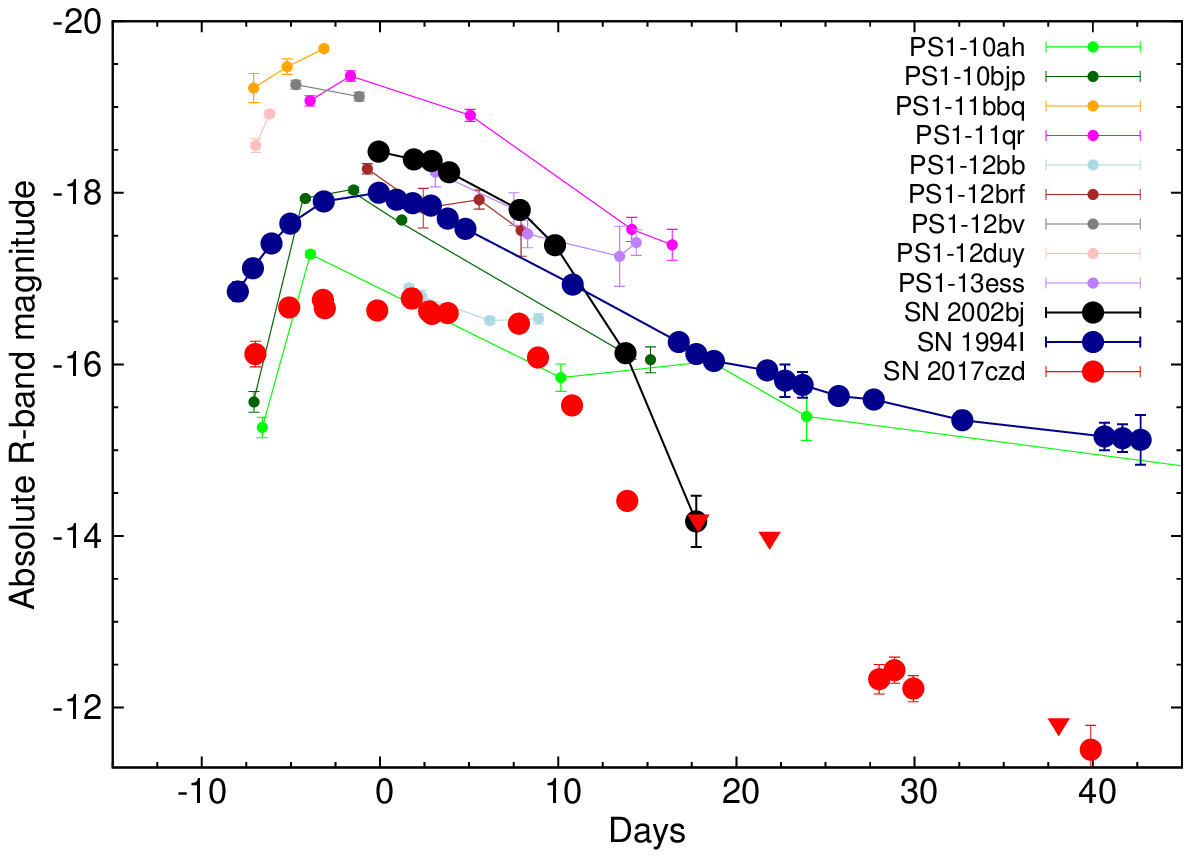}
\includegraphics[width=8cm,clip]{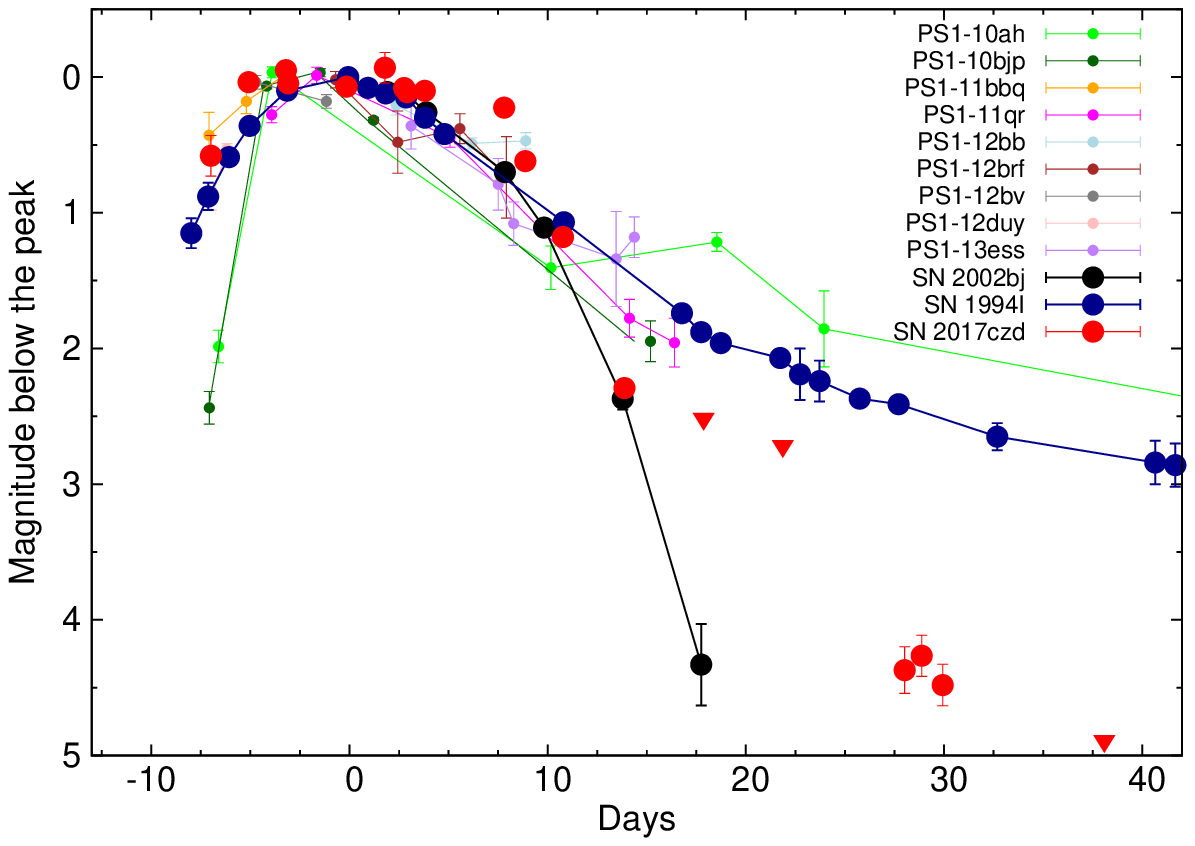}
\caption{(Top) The $R$-band absolute magnitudes of SN~2017czd compared with 
those of rapidly evolving transients by \citet{drout2014} in the $r$-band.
The extinction of each SNe has been corrected.
The days of these SNe are shifted to match their maximum dates. 
(Bottom) The same as the top panel but the magnitudes  relative to the peaks.
Both figures are compared in the rest frames.}
\label{fig:abs_fast}
\end{figure}

\begin{figure}[t]
\centering
\includegraphics[width=8cm,clip]{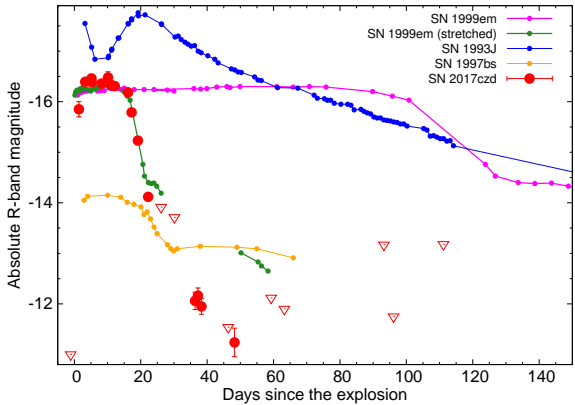}
\caption{The $R$-band absolute magnitudes of SN~2017czd compared with
those of SNe IIb 1993J, IIn 1997bs, and IIP 1999em. 
The green curve shows the light curve of SN 1999em 
whose timescale is stretched by a factor of 0.17.
}
\label{fig:com_sc}
\end{figure}

\begin{figure}[t]
\centering
\includegraphics[width=8cm,clip]{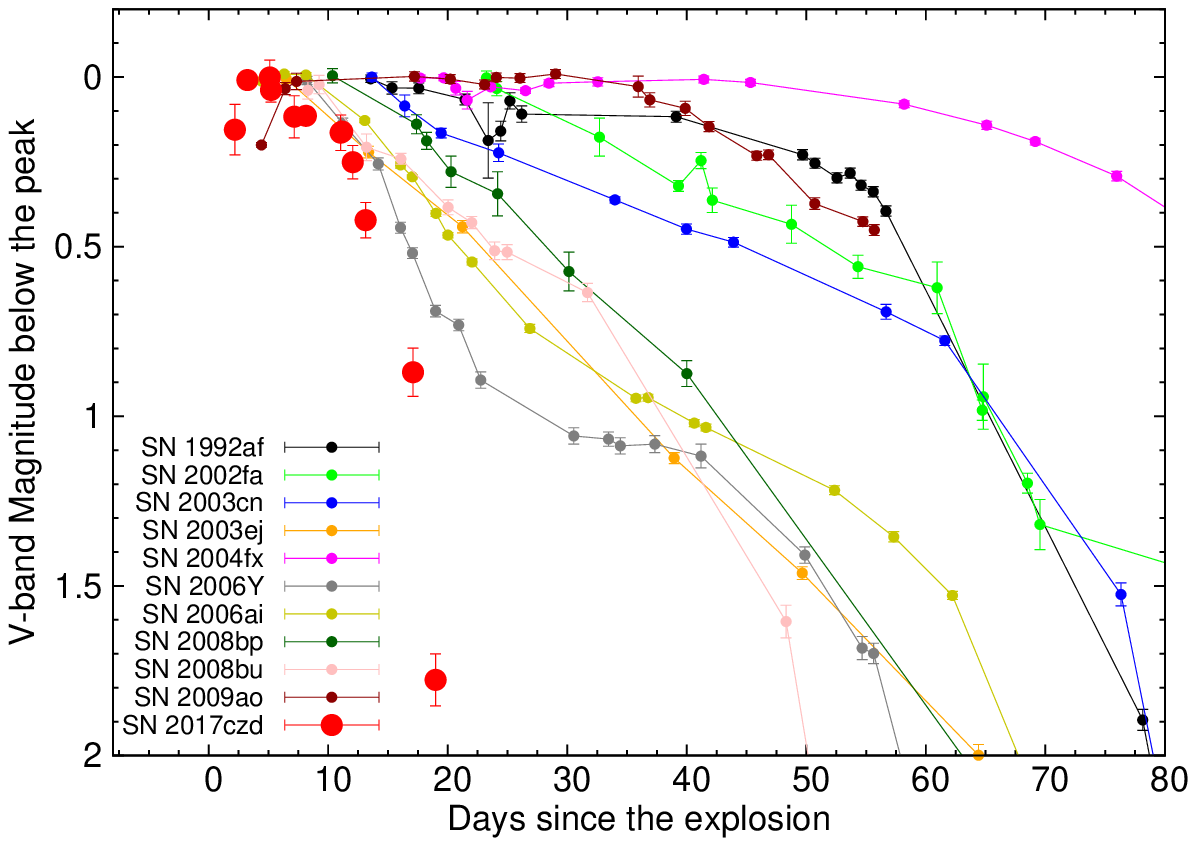}
\caption{
The $V$-band light curves of SN~2017czd and SNe~II with a short plateau in \citet{anderson2014}.
}
\label{fig:short}
\end{figure}

The light curves of SN 2017czd can be divided into four stages based on the slopes of the $R$-band light curve:
(i) the rising phase (up to $t \sim \hs 3$~days),
(ii) the plateau phase (between $t \sim \hs 3$ and 16~days),
(iii) the declining phase (between $t \sim \hs 16$ and 30~days),
and (iv) the tail phase (after $t \sim \hs 30$~days).
Below we discuss the light-curve properties in each stage.

A rapid rise is found at the early phases.
The rising rate is $\sim \hs 0.3$~\magdays in the $R$ band between $t=$ 1.4 and 3.1~days.
Using the $i$ band magnitude of 17.8 mag obtained by the Pan-STARRS at $t=$ 1.1~days\footnote{\url{https://wis-tns.weizmann.ac.il/object/2017czd}}, we obtain the rising rate of $\sim \hs 1.6$~\magdays in the $I$ band between $t=$ 1.1 and 2.1~days.
Combined with the fact that the pre-discovery upper-limit magnitude of 21.5 mag at $t=$ -1.2 days, it is evident that this SN experiences a rapid rise just after the explosion.

Between $t \sim 3$ and 16~days, the light curves show a short plateau except for the $B$ band.
The flat shape of the light curves are most evident in the $R$ and $I$ bands.
We define $t_{p}$ as the duration in which magnitude is constant within 0.2 mag. We find $t_{p} \sim \hs 13$~days in $R$ and $I$ bands.
In the $V$ band, the plateau is also seen, but the duration is  shorter than that in the $R$ and $I$ bands, i.e.,
the $V$-band light curve starts to decline around $t=$ 13~days. 
In the $B$ band, the light curve reaches the peak at $t \sim \hs 5$~days and then starts to decline.
These characteristics of the plateau is similar to the normal SNe IIP except for the plateau length (see also Figure \ref{fig:com_sc}).
Given the sparse NIR data, the plateau shapes in the $J$ and $H$ band light curves are not as clear as in the $R$ and $I$ bands, but they also show relatively flat light curves between $t \sim \hs 3$~and 16~days.

After the plateau, optical light curves show the rapid declines.
In particular, the $R$ and $I$ band light curves suddenly drop after their plateau phases.
The light curves in shorter wavelengths exhibit more rapid evolution: the decline rates between $t=15.9$~days and 21.9~days are estimated to be 0.5, 0.3, and 0.2~\magdays
in the $V$, $R$, and $I$ bands, respectively. 
We see a {slowdown in the decline rate of the $R$ band light curve after $t \sim 35$~days (0.1~\magdays) although the photometric error at $t=47.7$~days is rather large.

\subsection{Comparison with Other Transients}
Compared with normal SNe, the light curves of SN 2017czd show more rapid evolution,
but its absolute magnitudes are within the range of normal SNe.
\red{Figure \ref{fig:abs_normal} compares the $R$-band magnitude of SN~2017czd with those of normal SNe; SNe Ib 2008D \citep{modjaz2009},
IIL 1986L \citep{anderson2014}, IIb 1993J \citep{richmond1996_93J},
IIP 1999em \citep{leonard2002_99em}, IIb 2008ax \citep{pastorello2008}, and IIn 2009ip \citep{2013MNRAS.433.1312F}}.
In order to compare their rise times quantitatively, we define the time it takes from the half maximum luminosity to reach the maximum luminosity as $t_{1/2,\mathrm{rise}}$. 
For SN 2017czd,
it is estimated to be 2.6~days, which is much shorter than those of
SNe 1993J (7.64~days) and 2008ax (9.05~days). 

The $R$-band magnitude around the peak is $-16.5$ mag for SN 2017czd,
which is comparable to those of SNe 1999em and 2008D.
\red{After the light-curve drop, the absolute magnitude of SN 2017czd 
becomes much fainter than those of SNe 1986L, 1993J, 1999em, 
2008D, 2008ax, and 2009ip.}
The luminosity at the tail phase becomes comparable to faint SN Iax 2008ha \citep{foley2009, valenti2009}.

To compare the decline rates after the plateau, 
the bottom panel of Figure \ref{fig:abs_normal} shows the 
magnitude evolution relative to the peak.
The decline rate of SN~2017czd (0.3 \magdays in $R$ band) is the highest among other SNe compared in Figure \ref{fig:abs_normal}.
For example, the magnitude decline by $4.5$ mag in $\sim 20~\mathrm{days}$ after the plateau for SN 2017czd, while it is 1.5 mag for SN 2008ha.

The fast decline of SN 2017czd is in fact similar to those of rapidly evolving transients.
Figure \ref{fig:abs_fast} shows the $R$-band light curve 
of SN~2017czd compared with $r$-band light curve of the rapidly evolving transients by \citet{drout2014},
and $R$-band light curve of the fast declining SN Ic 1994I \citep{richmond1996_94I} and SN 2002bj \citep{poznanski2010}.
A large diversity of $\sim \hs 3$ mag is seen in the peak absolute magnitudes in the rapidly evolving transients of \citet{drout2014}, 
and SN 2017czd is located at the faint end.
Among rapidly evolving transients presented by \citet{drout2014}, 
the observations after $t \sim \hs 20$~days after the maximum are available only for PS1-10ah, and SN 2017czd shows a faster decline than PS1-10ah.

The bottom panel of Figure \ref{fig:abs_fast} compares the light curves relative to the peak. 
Compared with other rapidly evolving transients in the figure, only SN 2017czd exhibits the plateau in the light curve.
Because the data are sparse for the rapidly evolving transients,
we define the decline time ($t_{1/2,\mathrm{decline}}$) as the time it takes from the light-curve maximum to the half maximum for the comparison. 
The timescale of SN 2017czd ($t_{1/2,\mathrm{decline}} =14.5$~days in the $R$-band) is similar to those of SN 2002bj (8.3~days in the $R$-band, \citealt{poznanski2010}) and the Pan-STARRS rapidly evolving transients ($t_{1/2,\mathrm{decline}} =$3--17~days in the $r$-band, \citealt{drout2014}).

Motivated by the presence of the short plateau in the light curves,
we also compare the light curves of SN 2017czd with SNe II.
Figure \ref{fig:com_sc} shows the $R$-band light curve of SN~2017czd compared with SNe 1993J, 1999em, and SN impostor 1997bs \citep{vandyk2000}.
As in other Type IIP SNe, SN 1999em shows a plateau for $\sim$100~days in the $R$-band.
We stretch the $R$-band light curve of SN IIP 1999em to match that of SN 2017czd by assuming a stretch factor of 0.17 in time (green in Figure \ref{fig:com_sc}).
The stretched light curve is in fact quite similar to that of SN 2017czd until $t\sim \hs 20$~days. 
The sharp drop after the plateau is also similar, although the light curve of SN 2017czd keeps declining rapidly.

\red{
Some SNe~II are known to show a fast evolving light curves
\citep[so called SNe IIL, e.g.,][]{faran2014_IIL,bose2014,bose2016}.
Figure \ref{fig:short} shows the $V$-band light curve of SN~2017czd
and those of such SNe~II presented in \citet{anderson2014}.
The $V$-band light curve of SN~2017czd until 15~days 
has a similar evolution to that of SN~2006Y, which shows the fastest decline among their samples, 
while SN 2017czd declines more rapidly after 15~days.
As shown in Figure \ref{fig:short},
the duration of the plateau of SN 2017czd is among the shortest 
in the samples of SNe II collected by \citet{anderson2014}.
}

It is interesting to note that
SN (or SN impostor) 1997bs also shows a short plateau in the $R$-band light curve (Figure \ref{fig:com_sc}), whose timescale is  comparable to that in SN 2017czd.
However, several characteristics are different from those of SN~2017czd:
the plateau luminosity of SN 1997bs is lower than that of SN 2017czd, and
the light curve after the short plateau becomes flat again.
Furthermore, spectra of SN 1997bs show narrow emission lines,
which are not seen in SN 2017czd (see Section \ref{spec}).

Figure \ref{fig:color} is compared with the $V-R$ color evolution of SN~2017czd and some Type II SNe; SNe IIb 1993J \citep{richmond1996_93J},
IIP 1999em \citep{leonard2002_99em}, and IIb 2008ax \citep{pastorello2008}.
Until $t \sim \hs 10$~days, the $V-R$ color of SN~2017czd is similar to or slightly redder than those of SNe 1993J and 1999em.
However, after $t \sim \hs 15$~days, the color in SN 2017czd rapidly becomes  redder.
Such a rapid change is not seen in other SNe IIb and IIP,
confirming that SN 2017czd is a rapidly evolving object also in color.

\begin{figure}[t]
\centering
\includegraphics[width=8cm,clip]{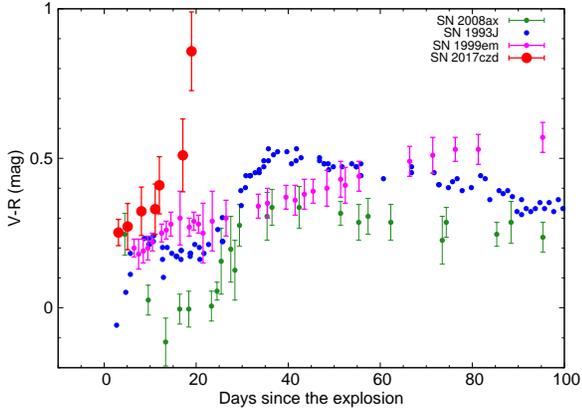}
\caption{The $V-R$ color evolution of SN~2017czd compared with those of SNe IIb 1993J, 2008ax, and SN IIP 1999em.
The extinction of each SNe has been corrected.}
\label{fig:color}
\end{figure}


\section{Spectra}\label{spec}
Figure \ref{fig:spec} shows the time evolution of the optical spectra
from the rising phase ($t=2.1$~days) to the declining phase ($t=19.0$~days).
Note that narrow emission lines of \Ha, \Hb, [O~\siii] $\lambda \lambda$ 4959, 5007, [N~\sii] $\lambda$ 6583, and [S~\sii] $\lambda \lambda$ 6716, 6731 come from the H II region in the host galaxy.
Initially, the spectra are dominated by the blue continuum as in SNe IIP in their early phases \citep[e.g.,][]{huang2018}.
The broad absorption component of \Ha around 6000~\AA\ is present from $t=3.1$~days.
This broad \Ha feature led us to classify SN~2017czd as an SN~II as discussed below.
In addition to \Ha, the broad feature of Ca~\sii IR triplet are seen in the spectra at $t=16.0$ and 19.0~days.

The spectra of SN 2017czd are similar to those of SNe IIP in early phases.
Figure \ref{fig:com_sp} shows the spectral comparison with various types of SNe; 
SN~2017czd at $t=7$~days, SN IIP 2006bp at $t=6$~days \citep{quimby2007},
\red{SN~II~2006Y at $t=11$~days \citep{guti2017},}
SN IIb 1993J at $t=5$~days \citep{barbon1995}, 
SN IIn 1998S at $t=2$~days \citep{fassia2001}, luminous SN Ia 1991T at $t=10$~days \citep{mazzali1995}, and broad-lined SN Ic 2006aj at $t=14$~days \citep{pian2006}.
The blue continuum and the broad \Ha line in the spectrum of SN~2017czd match those of SN~2006bp at $t=6$~days. 
This supports our classification of SN~2017czd as a SN II.
SNe~1993J and 2006aj show more prominent absorption lines than SN~2017czd.
\red{SNe 1998S, 2006Y and 1991T also show a blue continuum.
However, SN 1998S shows characteristic strong narrow emission lines observed in SNe~IIn,
SN~2006Y has no features except for the \Ha emission from the host galaxy and
SN 1991T shows a strong absorption feature of  Fe~\siii around 5000 \AA.}
Thus, they are clearly different from SN 2017czd.

In Figure~\ref{fig:com_sp5}, we compare SN~2017czd with type II SNe,
SNe IIP 1999em \citep{leonard2002_99em}, 2012aw \citep{bose2014}, and SNe IIb 1993J \citep{barbon1995}, 2008ax \citep{pastorello2008} at around $t=5$~days.
The \Ha absorption line of SN 2017czd is shallow and blueshifted compared with those of SNe IIP.
Figure~\ref{fig:com_sp20} shows the comparison with the same SNe but at the later epochs.
The spectrum of SN~2017czd shows weak absorption lines of \Ha, Ca~\sii IR triplet,
and He~\si $\lambda$ 5876 at $t=20$~days.
The spectral features of SN~2017czd are broadly similar to SNe IIP at early phase ($\sim \hs 5$~days) and similar to SNe IIb SN~1993J at decline phase ($\sim \hs 20$~days).

Figure \ref{fig:ha} shows the evolution of the \Ha line velocity in SN~2017czd.
We fit the Gaussian function to the \Ha absorption line profile.
We consider the difference between the rest wavelength and the wavelength at the minimum of the fitted Gaussian as the line velocity.
The \Ha velocity of SN 2017czd is about $\sim$26,000~\kms at $t \sim \hs 3$~days and declines to $\sim$20,000~\kms at $t\sim$15~days.
These velocities are significantly higher than those of other SNe II at similar epochs,
which are around 10,000--20,000~\kms.

\begin{figure}[h]
\centering
\includegraphics[width=8cm,clip]{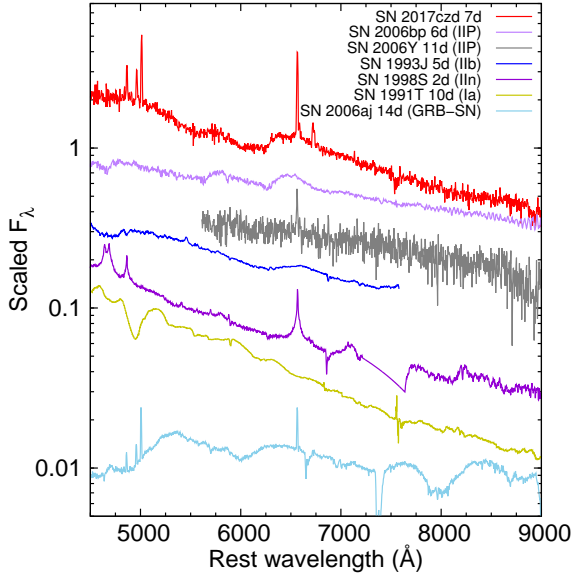}
\caption{Spectrum of SN~2017czd at 7~days compared with well-observed SNe IIP 2006bp, IIP 2006Y, IIb 1993J, IIn 1998S, Ia 1991T, and 
broad-lined Ic 2006aj at similar epochs. The fluxes were shown in the logarithmic scale and arbitrarily scaled to avoid overlaps.
}
\label{fig:com_sp}
\end{figure}

\begin{figure}[h]
\centering
\includegraphics[width=8cm,clip]{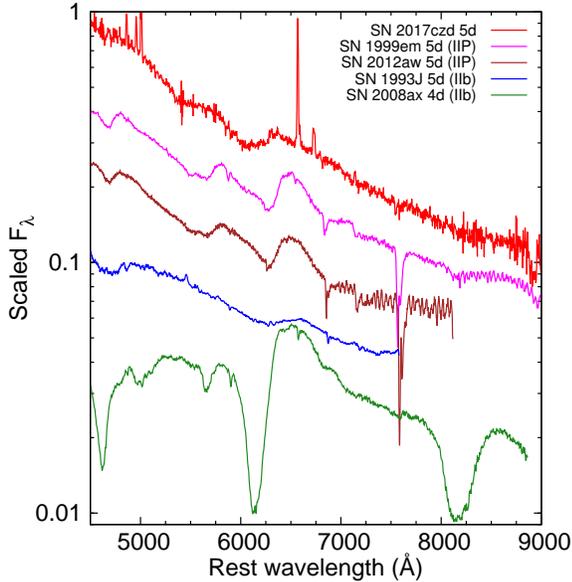}
\caption{
Spectrum of SN~2017czd at $t \sim 5$~days compared with those of hydrogen-rich SNe at the same epoch.
}
\label{fig:com_sp5}
\end{figure}

\begin{figure}[h]
\centering
\includegraphics[width=8cm,clip]{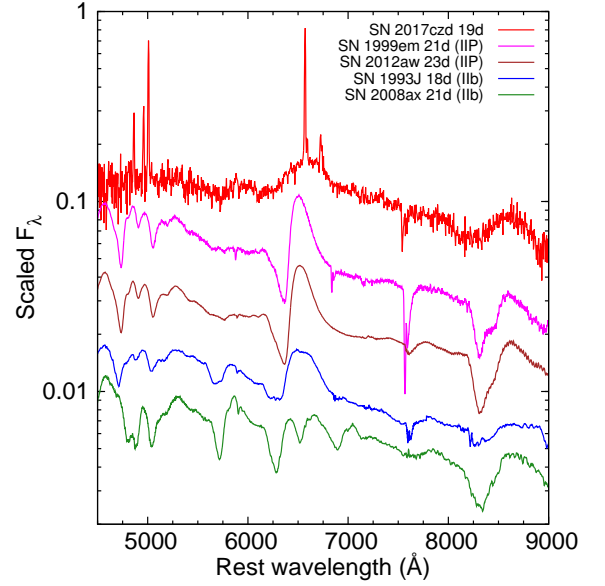}
\caption{
The same as Fig.~\ref{fig:com_sp5} but at around $t \sim 20$~days.
}
\label{fig:com_sp20}
\end{figure}

\begin{figure}[h]
\centering
\includegraphics[width=8cm,clip]{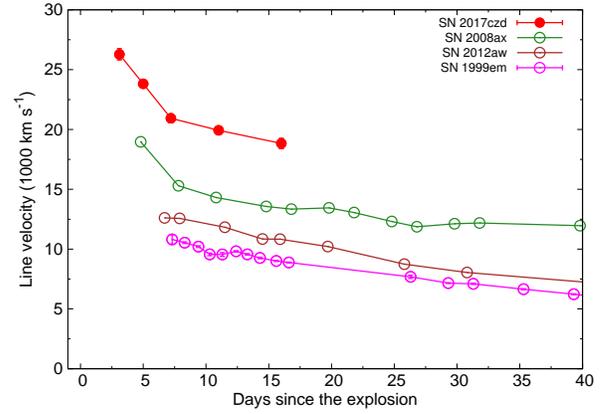}
\caption{\Ha velocity evolution of SN~2017czd compared with those of SNe IIP 1999em, 2012aw, and IIb 2008ax. The errors are from the wavelength calibration and fitting uncertainties.
The wavelength resolution corresponds to the velocity resolution of 750~\kms.
}
\label{fig:ha}
\end{figure}


\section{Discussion}
We have shown that SN~2017czd has different light curve characteristics from those of other rapidly evolving transients currently known and it shows a plateau phase as in SNe~IIP although it is very short (Section~\ref{sec:lightcurve}). We have also presented that SN~2017czd clearly shows the broad hydrogen feature in the spectra (Section~\ref{spec}). Based on these facts, we discuss the nature of SN~2017czd in this section.

\subsection{Power source}\label{sec:powersource}
First, to clarify the power source of SN~2017czd, we construct the bolometric light curve based on our optical and infrared photometry. We find that the spectral energy distribution (SED) in the $BVRIJH$-band
can be well explained by a single black-body function until the end of the plateau phase when the multi-band photometry is available. Therefore, we estimate the bolometric luminosity of SN 2017czd from the rising phase
to the declining phase by integrating the single black-body function matching the photometry.

Because we only have the $R$-band photometry during the tail phase after the plateau, we estimate the bolometric luminosity in the tail phase based only on the $R$-band photometry.
\citet{lyman2014} investigate the bolometric correction of SNe IIP and SNe IIb
and find that the $R$-band should covers 20--25\% of the total fluxes in the tail phases.
Assuming this fraction (20\%), we obtain the quasi-bolometric luminosity of SN 2017czd at the tail phase.
The bolometric light curve of SN 2017czd is presented in Figure \ref{fig:bol}.

One possible power source of SN 2017czd is the radioactive decay of $^{56}$Ni, which is a standard power source of SNe. The total radioactive luminosity from the decay of \Nifs is \citep{nadyozhin1994}
\begin{eqnarray}
\left[6.5\exp\left(\frac{-t}{8.8~\mathrm{days}}\right) + 1.45\exp\left(\frac{-t}{111.3~\mathrm{days}}\right)\right] \frac{M_{{}^{56}\mathrm{Ni}}}{M_{\odot}} 10^{43}\ \mathrm{erg~s^{-1}},\nonumber\\
\label{eq:str}
\end{eqnarray}
where $M_{{}^{56}\mathrm{Ni}}$ is the initial mass of radioactive \Nifs.

Assuming that the major power source at the tail phase is the \Nifs decay, 
we compare the estimated bolometric light curve with the total available decay energy (Eq.~\ref{eq:str}) in Fig.~\ref{fig:bol}. The total luminosity is consistent with the total decay energy from 0.005~\Msun\ of $^{56}$Ni.
However, the peak luminosity of SN~2017czd is much brighter than the expected luminosity from this $^{56}$Ni mass assuming the full trapping of gamma-rays from the \Nifs decay. 
\red{Therefore, radioactive decay of $^{56}$Ni is unlikely as a power source for SN 2017czd.}

\red{
Another possibility is that the light curve is powered by the interaction between SN ejecta and CSM.
In fact, some SNe~IIn are known to show a plateau phase in the light curves \citep[e.g.,][]{2004MNRAS.352.1213C,2012MNRAS.424..855K,2013MNRAS.431.2599M}.
But their plateau phase lasts more than 100~days,
they are more luminous than SN~2017czd and they have strong hydrogen emission.
In addition, interacting the SNe with the CSM are usually observed as SNe~IIn with strong hydrogen emission lines, which are not observed in SN 2017czd.
Therefore, although the possible contribution from the interaction between SN ejecta and CSM to the plateau part is not clear, 
we believe that the interaction is unlikely to be a main power source.
}

\begin{figure}[h]
\centering
\includegraphics[width=8cm,clip]{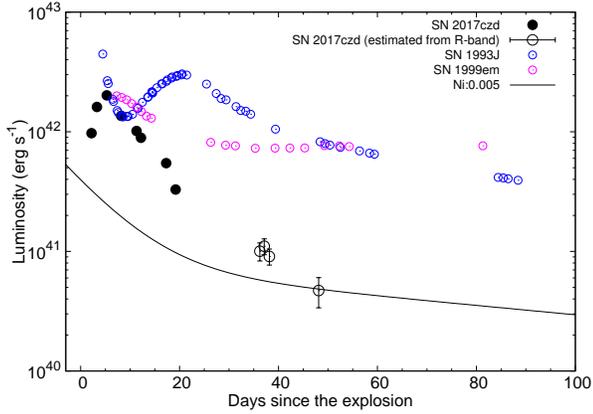}
\caption{Bolometric light curve of SN~2017czd compared with those of SNe IIP 1999em \citep{leonard2002_99em} and IIb 1993J \citep{lewis1994}.
The total available energy from the nuclear decay of $0.005~\Msun$ of \Nifs is presented.
}
\label{fig:bol}
\end{figure}

\subsection{SN 2017czd as a weak explosion of a SN IIb progenitor}
\label{nature}

In this section, we argue that the luminosity source of the early phase of SN~2017czd is the thermal energy provided by the SN shock as in the case of SNe~IIP during the plateau phase.

Assuming that SN~2017czd is powered by the thermal energy from the SN shock as in SNe IIP, we first estimate
the SN properties to account for the observed properties of SN~2017czd by using the scaling relation formulated by \citet{popov1993}.
We adopt $-16.8$~mag as the absolute $V$-band magnitude during the plateau and the plateau length $t_p = 13$~days.
Fe~\sii line velocity is typically used to estimate the photospheric 
velocity in SNe IIP but Fe~\sii absorption lines are not found in the spectra of SN~2017czd.
We only obtain the \Ha line velocity ($\sim 20000$ \kms at $t= 16.0$~days).
Fe~\sii line velocities for SNe IIP are roughly a half of the \Ha \citep[e.g.,][]{bose2014}. Assuming this relation, we set the photospheric velocity of $v_{ph}=10000$~\kms. With these parameters, the explosion energy is estimated to be $3\times 10^{49}~\mathrm{erg}$. The hydrogen-rich envelope mass and radius of the progenitor of SN~2017czd are estimated to be 0.04~\Msun\ and 860~\Rsun, respectively.

Such an extended progenitor with a small amount of the hydrogen-rich envelope is similar to the progenitors of SNe IIb which result from the evolution of the massive binary stars \citep[e.g.,][]{2012ApJ...757...31B,2017ApJ...840...90O,2017ApJ...840...10Y}. To confirm if such a progenitor can explain the observational properties of SN~2017czd, we perform numerical light curve calculations using a progenitor model obtained from stellar evolution modeling.

We take a progenitor model with a small hydrogen-rich envelope calculated by \citet{2017ApJ...840...90O} using \texttt{MESA} \citep{2011ApJS..192....3P,2013ApJS..208....4P,2015ApJS..220...15P,2016ApJS..223...18P,2018ApJS..234...34P}. The progenitor we adopt has $16~M_\odot$ and solar metallicity at the zero-age main sequence (ZAMS). It is in a binary system in which the initial secondary star is $15.2~M_\odot$ (the mass ratio of 0.95) and the initial period is 1000~days. The mass transfer in the binary system is treated in the non-conservative way with the mass accretion efficiency of 0.5. We refer to \citet{2017ApJ...840...90O} for further details of the progenitor evolution. The progenitor evolution is followed until the end of the central carbon burning. The envelope structure is not much affected by the rest of the evolution and we take this model for the light curve modeling. The final progenitor mass is $5.4~M_\odot$ with the hydrogen-rich envelope mass of $0.4~M_\odot$ and the helium core mass of $5.0~M_\odot$. The hydrogen fraction in the envelope is 0.46. The progenitor radius is $767~R_\odot$. The progenitor properties match to those of extended SN IIb progenitors \citep[e.g.,][]{2010ApJ...711L..40C}.

The synthetic light curves are calculated by using \texttt{STELLA}, which is a one-dimensional multi-group radiation hydrodynamics code developed by \citet{1998ApJ...496..454B,2000ApJ...532.1132B,2006A&A...453..229B}. We set the mass cut of the progenitor at $1.4~M_\odot$ and put the thermal energy just above the mass cut to initiate the SN explosions. We also put the radioactive $^{56}$Ni just above the mass cut.

Figure~\ref{fig:lcmodel_0p003} presents the synthetic multi-color light curves. The explosion energy and $^{56}$Ni mass of the model is $5\times 10^{50}~\mathrm{erg}$ and $0.003~M_\odot$, respectively. 
\red{The synthetic light curves after $\sim$ 35~days follow the decline rate of the light curve expected from the $^{56}$Co decay.}
Both the multi-color and bolometric light curves are well reproduced by this model. Given the constraint from the Gaia upper limit, SN~2017czd is found to rise more quickly than our model but the early rise may be caused by the confined dense CSM around the progenitor \citep[e.g.,][]{moriya2017,moriya2018,forster2018,2018ApJ...858...15M}.
The early discrepancy in the bolometric light curves is due to the fact that the bolometric luminosity constructed from the observations does not include contributions from the ultraviolet wavelengths which are included in the model.
While the model shows some discrepancy in the tail phase in the bolometric light curve,
we note that the bolometric luminosities in the tail has a large uncertainty,
since it is constructed solely by the $R$ band.
\red{
The photospheric velocity of the model presented in Figure~\ref{fig:lcmodel_0p003} is consistent with that estimated from the observation ($v_{ph} = 10000~\mathrm{km~s^{-1}}$), although our model shows slightly different time evolution. This difference might be due to our use of the Rosseland-mean opacity in estimating the photospheric velocity.
}
The estimated explosion energy and hydrogen-rich envelope mass by our numerical modeling are about a factor of 10 more than those estimated by the Popov formula. We presume that this is partly because of the large helium fraction in our envelope \citep{kasen2009}, as well as the limitation of the simple semi-analytic formula.

Figure~\ref{fig:lcmodel_nickelvariation} shows the synthetic light curves with the same progenitor models and explosion energy but with different $^{56}$Ni masses. The synthetic light curve with the large amount of $^{56}$Ni ($0.1~M_\odot$) shows the secondary peak due to the $^{56}$Ni heating as in ordinary SNe IIb.


Overall, the observations of SN~2017czd are well explained by the explosion of the SN~IIb progenitor.
The \Nifs mass of SN~2017czd ($\sim 0.001~\Msun$) is much smaller than those of SNe IIb ($\sim 0.1~\Msun$, e.g., \citealt{lyman2016}).
Thus, we suggest that SN~2017czd is a weak explosion from the SN IIb progenitor.
This explosion would not eject much $^{56}$Ni and the secondary luminosity peak due to the nuclear energy input would not be observed \citep[see also][]{milisav2013}.
It is worth noting that such a weak explosion with a small amount of $^{56}$Ni may be consistent with the explosions obtained by the current state-of-the-art neutrino-driven core-collapse SN simulations. It is well-known that the current core-collapse SN simulations fail to reproduce the standard SN explosions with the explosion energy of $\sim 10^{51}~\mathrm{erg}$ \citep[e.g.,][]{2016MNRAS.461L.112T}. In addition, \citet{2017arXiv170404780S} recently point out that the amount of $^{56}$Ni produced in the explosions obtained by the current numerical simulations would be very small because they take too much time for the shock recovery. Although the current simulations are yet far from reproducing the ordinary core-collapse SN explosions, their results match the explosion properties of SN~2017czd.
\red{
Alternatively, the low \Nifs mass may also be a result of fallback \citep{2007ApJ...660..516T,moriya2010}. A weak explosion can result in the unsuccessful explosion of the inner layers of the progenitor where \Nifs exists. Then, little \Nifs remains in the ejecta with a small explosion energy.
}

Our result indicates that some rapidly evolving transients may be related to the weak explosions of SNe IIb progenitors with little $^{56}$Ni ejection.
The $^{56}$Ni-free core-collapse SN explosions to account for rapidly evolving transients have been previously suggested for hydrogen-free extended progenitors \citep[e.g.,][]{2014MNRAS.438..318K,2018MNRAS.tmpL.182K}. However, we propose that there is a little hydrogen left in the progenitor in the case of SN~2017czd.
This makes the extremely short plateau in the light curve,
and the hydrogen features are also expected to be observed in this case.

\begin{figure}[h]
\centering
\includegraphics[width=\columnwidth]{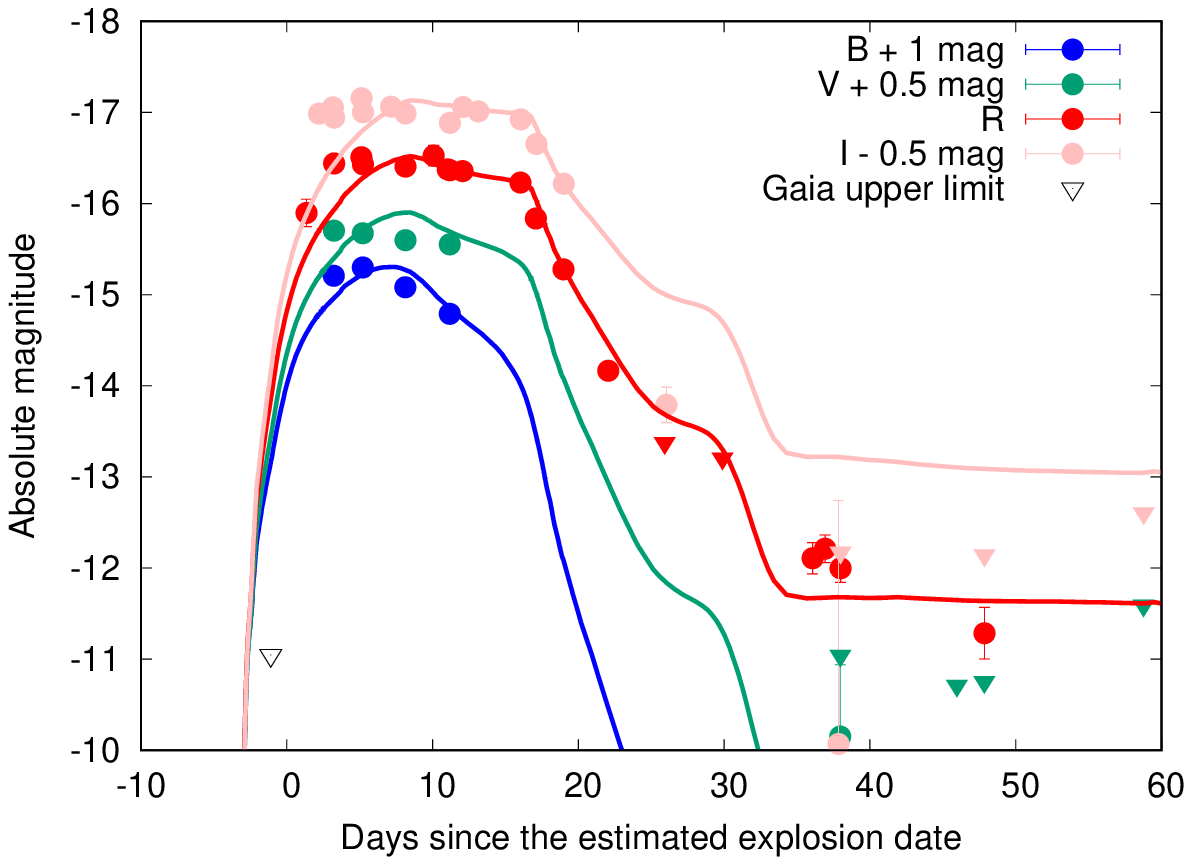}
\includegraphics[width=\columnwidth]{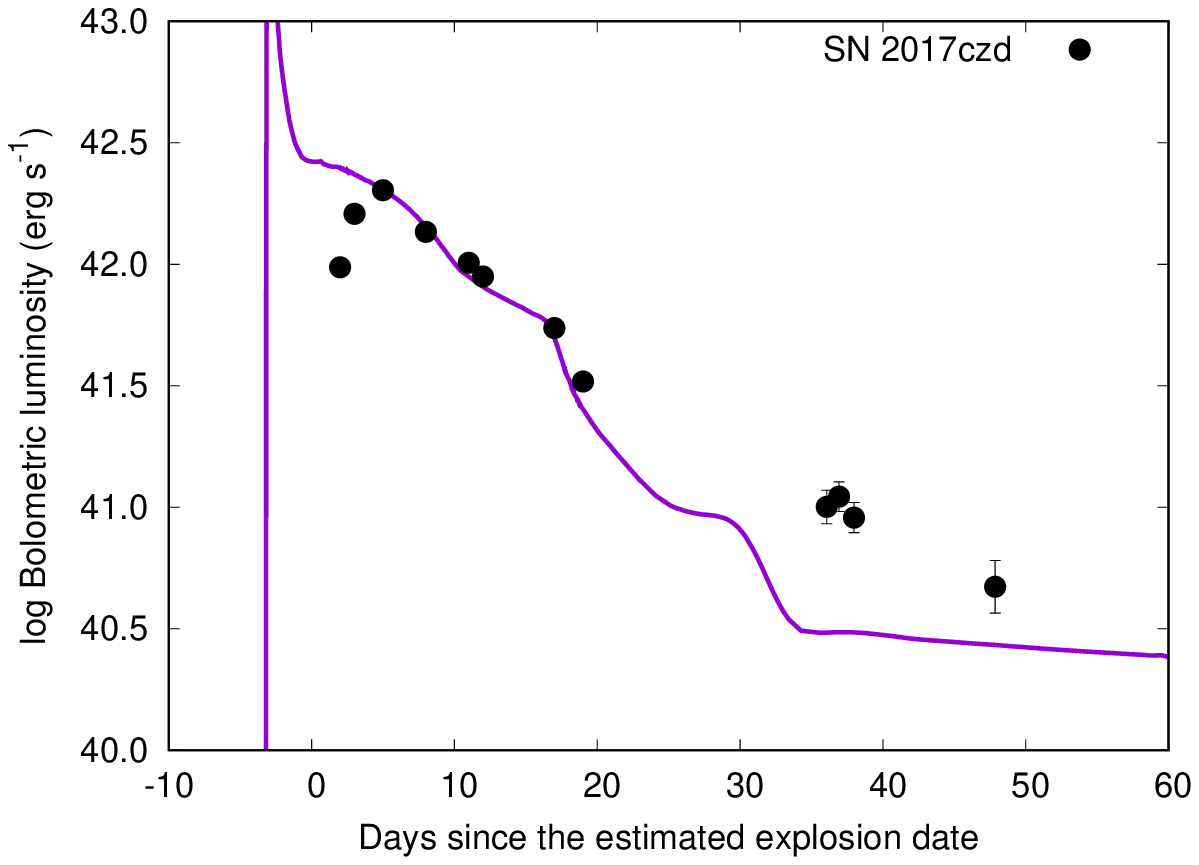}
\includegraphics[width=\columnwidth]{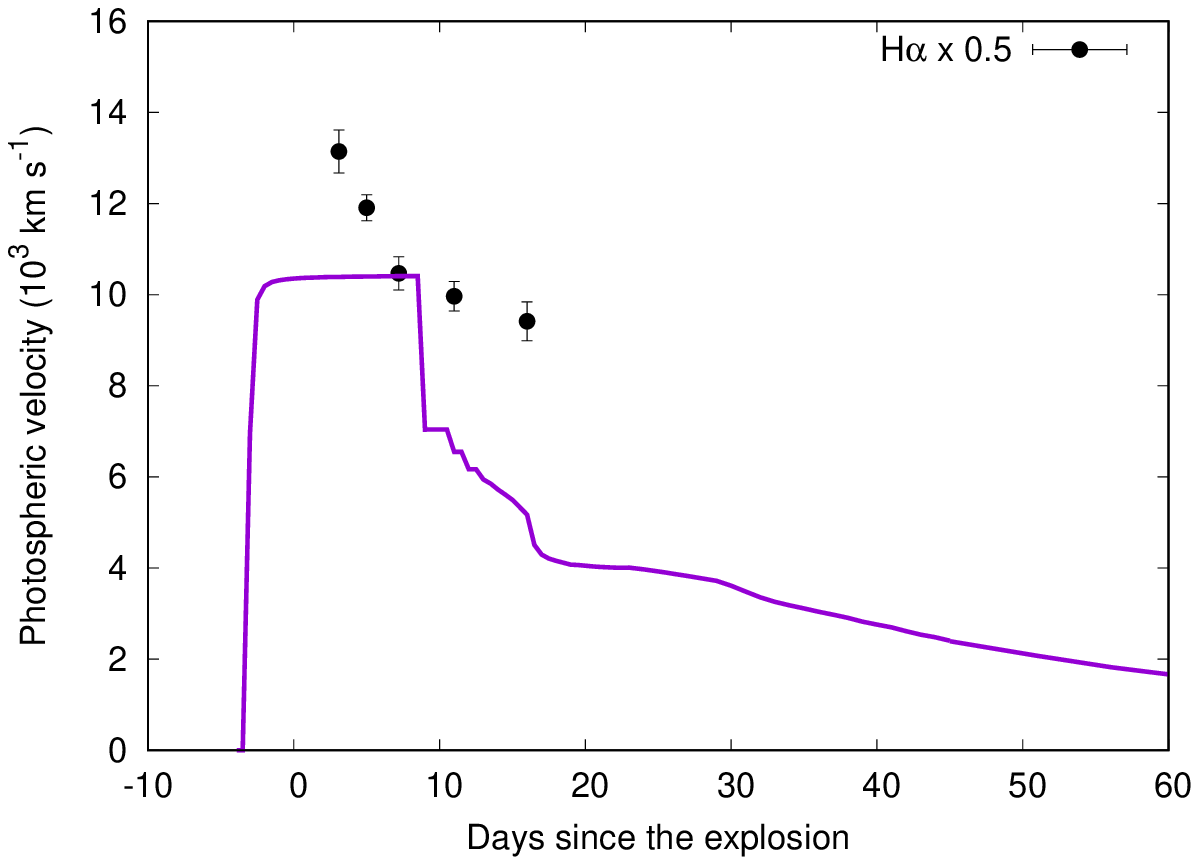}
\caption{
Synthetic multi-color (top) and bolometric (middle) light curves from our $5.4~M_\odot$ progenitor model with the hydrogen-rich envelope of $0.4~M_\odot$. The explosion energy and $^{56}$Ni mass of the model are $5\times 10^{50}~\mathrm{erg}$ and $0.003~M_\odot$, respectively. The open triangle shows the Gaia upper limit. The bottom panel shows the photospheric velocity evolution of the model, where the photosphere is defined as the radius with the Rosseland-mean optical depth of 2/3. \red{The half of the \Ha velocity (Fig.~\ref{fig:ha}) which approximately traces the photospheric velocity is also shown.} The time in the figure is from the observationally estimated explosion date and the explosion date of our synthetic model is 4~days before the estimated explosion date.
}
\label{fig:lcmodel_0p003}
\end{figure}

\begin{figure}[h]
\centering
\includegraphics[width=\columnwidth]{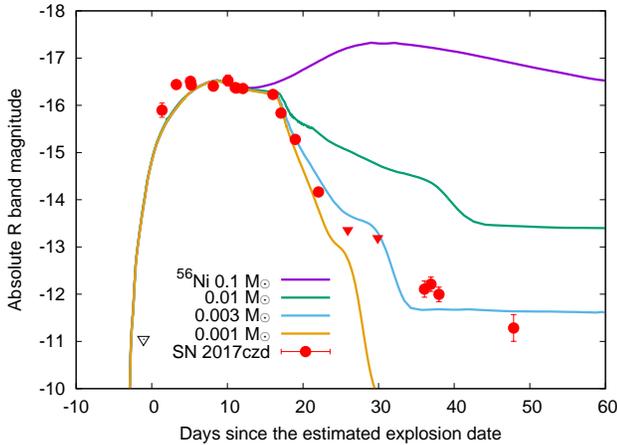}
\caption{
The $R$-band synthetic light curves with the same progenitor and explosion energy as in Fig.~\ref{fig:lcmodel_0p003} but with different $^{56}$Ni mass. The open triangle shows the Gaia upper limit.
}
\label{fig:lcmodel_nickelvariation}
\end{figure}


\section{Summary}
We present our optical and NIR observations of rapidly evolving SN 2017czd with hydrogen features.
The light curves of SN 2017czd show a short plateau ($\sim$13~days in $R$ band) followed by a rapid decline.
The decline rate of SN 2017czd (0.3~\magdays in $R$ band) is faster than the standard SNe IIP and IIb (Fig. \ref{fig:abs_normal}), while it is similar to
those of the rapidly evolving transients \citep[\eg][ Figure~\ref{fig:abs_fast}]{drout2014}.
The peak absolute magnitude ($-16.5$~mag in $R$-band) is consistent with those of SNe IIP and IIb.
The peak luminosity is also consistent with those of rapidly evolving transients.
The spectra exhibit the hydrogen features, 
and overall spectra are similar to those of SNe IIP in the early phases and SNe IIb in the late phases. 
However, the hydrogen features are broader than those of SNe IIb and IIP,
and line velocities are larger.

We calculate synthetic light curves based on a binary progenitor model ($16~M_\odot$ at ZAMS and $5.4~M_\odot$ at the explosion) with a small  hydrogen-rich envelope ($0.4~M_\odot$) at the pre-explosion stage.
The observed properties of SN~2017czd, including the short plateau duration and rapid decline, are explained by the model with the relatively low explosion energy ($5 \times 10^{50}$~erg) and the low \Nifs mass ($0.003~\Msun$).
We conclude that SN~2017czd is a weak explosion of the SN IIb progenitor, which does not eject much $^{56}$Ni. 
Our results suggest that some rapidly evolving transients are also caused by such a weak explosion of SN~IIb progenitors.



\acknowledgments
This research has made use of the NASA/IPAC Extragalactic Database (NED) which is operated by the Jet Propulsion Laboratory, California Institute of Technology, under contract with the National Aeronautics and Space Administration. 

This research has made use of the NASA/ IPAC Infrared Science Archive, which is operated by the Jet Propulsion Laboratory, California Institute of Technology, under contract with the National Aeronautics and Space Administration.

Takashi J. Moriya is supported by the Grants-in-Aid for Scientific Research of the Japan Society for the Promotion of Science (JP17H02864, JP18K13585).

Masayuki Yamanaka is supported by the Grants-in-Aid for Young Scientists of the Japan Society for the Promotion of Science (JP17K14253).

Keiichi Maeda acknowledges support provided by Japan Society for the Promotion of Science (JSPS) through KAKENHI Grant JP17H02864, JP18H04585 and JP18H05223.

Sergei Blinnikov is supported by RSCF 18-12-00522 on the development of STELLA code.

Numerical computations were in part carried out on PC cluster at Center for Computational Astrophysics, National Astronomical Observatory of Japan.


\bibliographystyle{apj}
\bibliography{supernova}

\end{document}